\documentclass{appolb}

\begin{document}
\date{\today}
\pagestyle{plain}
\newcount\eLiNe\eLiNe=\inputlineno\advance\eLiNe by -1
\title{Heavy-Flavor Production Overview}
\author{Jeffrey A. Appel\\
Fermilab, PO Box 500, Batavia, IL 60510 USA\\
e-mail: appel@fnal.gov}
\maketitle

\begin{abstract}

This talk serves as an introduction to the Heavy-Flavor session of the 
XXXIII International Symposium on Multiparticle Dynamics.  A major focus 
of this session is on the production of heavy quarks.  The talks which 
follow review the latest results on heavy quark production in strong, 
electromagnetic, and weak interactions, as well as some of the physics of 
the heavy quarks themselves.  This talk emphasizes what we can learn from 
the production measurements, both about underlying QCD theory and the 
partonic nature of the hadrons which we see in the laboratory.

\end{abstract}

\section{Dedication}

We should start this session by recognizing the loss of Krzysztof Rybicki, 
until his death a member of the Local Organizing Committee and mover to 
have this session on heavy flavor reintroduced to the Multiparticle 
Dynamics program.  Krzysztof was the original organizer of this session, 
outlined its content, and personally invited me to participate.  We will 
all miss Krzysztof's presence here -- his intellectual contributions ... 
and his wonderful smile.  We dedicate this session to the memory of 
Krzysztof Rybicki.

\section{Introduction}

Today's program is filled with results from a large variety of physics
environments, involving production of heavy quarks in hadronic, 
electromagnetic, and weak interactions.  We will see a lot of data in the 
next six talks,[1-6] some of it quite new.  In what 
is presented, there are some themes of particular interest.  In the area 
of QCD Dynamics, there are results which test basic QCD theory and even 
what may lie beyond the Standard Model.  The data should help in 
navigating among recent theory improvements -- from 
next-to-next-to-leading-order and $k_t$-factorization calculations to 
various resummation techniques.  Also, there will be results which tell us 
about the structure of hadrons.  This information is more than just an 
input to QCD calculations, but fundamental information on the nature of 
and parameters describing quarks and gluons in hadrons.  By the end of the 
session, we should see what we have learned and what remains a 
mystery.\cite{mehen}

\section{Testing QCD}

Last month, Thomas Gehrmann, in his Lepton Photon 2003 review, 
''QCD Theoretical Developments", said that the ''testing QCD"-era has been 
over for some time.\cite{lp2003} I think that this is, perhaps, part of 
the general euphoria about the Standard Model.  Nevertheless, a quick 
search of SPIRES finds a significant number of papers with titles 
including ''test of" or ''testing" QCD.  The numbers are given in 
Table \ref{table1}.  
A significant fraction of the papers come from experiments with new 
results.  Testing is the job of experiments, after all; to probe the 
theory.  So, testing is not over yet!

\begin{table}
\caption{Recent SPIRES Listings on Tests and Testing of QCD}
\centering
\begin{tabular}{|c|c c c c c c|}
\hline
Year     & 1998 & 1999 & 2000 & 2001 & 2002 & 2003 \\
\hline
No. papers &   4  &  10  &  8   &  4   &  11  &  3 (so far) \\
\hline
\end{tabular}
\label{table1}
\end{table}

\subsection{Some Recent Surprises and Old Mysteries Still With Us}

We have had an unusual number of surprises this year involving heavy 
flavor production:

Double charm production rate in the $e^+e^-$ continuum

Production rate of double charm baryons by $\Sigma$'s

Better agreement with theory for charm than beauty production in 
\indent\indent$\gamma$-$\gamma$ collisions at LEP

\noindent
Older mysteries are still with us.  Consider the

$b$ quark production rates at Tevatron Collider energies

$J/\psi$ and $\psi'$ production rates at fixed target and collider
energies

\subsection{Heavy-Flavor Cross Sections}

The context for most thinking about heavy-flavor production depends on the 
factorization of the cross section, $\sigma_{QCD}$, into (1) a hard 
scattering of incident partons which have come from (2) distributions in 
the incident particles and (3) a hadronization process of the outgoing 
partons that result from the hard scattering.  The hard scattering is 
often drawn as a box.  We might describe the thinking in this context as 
thinking ''inside the box" -- to use an American colloquialism.

Typically, the measured cross sections are much larger than leading-order 
(LO) QCD predictions, even when renormalization and factorization scales 
and masses are set low.  Cross sections are still larger than 
next-to-leading-order (NLO) predictions, typically by factors 
$\sim$ 2, at least in some kinematic regions.  Recently, adding 
resummation effects (Next-to-Leading-Log, NLL, in $p_t$) and refined 
fragmentation functions are helping with agreement.  How should we 
interpret this progression?  Real progress?  Yes.  Furthermore, H. Jung, 
using a $k_t$-factorization calculation,\cite{jung} finds even better 
agreement with the data.  In this calculation, done with an appropriate 
set of parton distribution functions coming from HERA, some higher-order 
(resummed) terms are inherently included.  The technique also works 
surprisingly well for charm using the same parton distribution 
functions.\cite{jungprivate}

Perhaps it is worth noting that data is typically in a limited kinematic 
region.  An important step has just been made by CDF in triggering and 
analyzing B production down to $p_t$ of zero (using decays to $J/\psi$'s).  

One of the more disturbing things to note in the way we think about new 
theoretical calculations is our willingness to keep any Standard Model 
effect which increases the predicted $\sigma_{QCD}$.  This is hardly 
unbiased science.

\subsection{Charmonium Issues}

Measurements of $J/\psi$ production at fixed-target energies are a factor 
7 too large, and $\sigma_(\psi')$ is a factor 25 too large relative to 
the older leading-order calculations.\cite{e789J}  Can this be due to 
the color-octet mechanism in addition to the usual color-singlet 
mechanism?   Are color-octet matrix-elements as relevant at fixed-target 
energies as at the Tevatron Collider?  Actually, the color-octet 
parameters from the Tevatron don't work at HERA.  Furthermore, the 
polarization predicted at high $p_t$ for color-octet contributions has 
not been seen at the Tevatron.  What is going on here? Where is all the 
charmonium coming from?

Direct charmonium production is a small fraction of the total charm 
production.  Yet, color-octet hard scattering is a possible contribution, 
color evaporation, too.  ''Thinking outside the box" leads to the 
possibility of Non-Standard Model sources - e.g., light SUSY (see below).
We will need data over broader kinematic ranges to sort things out (e.g., 
lower $p_t$ where the cross section is largest).  Also, note the 
$p_t$-spectrum dependence of resummation in $p_t$.  This may be part of 
the answer to our questions.  Theory is only credible when terms are 
universal, non-process specific.  Yet, we have trouble, as noted, in 
relating color-octet contributions in hadronic interactions to 
electromagnetic interactions.  The direct charmonium production ''K 
factors" don't look like ''simple" higher order effects to me.

\subsection{Thinking Outside the Box}

This has all been thinking ''inside the box."  However, there is also 
''thinking outside the box."  Ed Berger and his colleagues have noted the 
possibility that new physics could account for an excess of $b$ 
production at the Tevatron.\cite{berger}  They assume the existence of a 
low-mass color-octet, spin 1/2 gluino and a low-mass color-triplet spin-0 
bottom squark.  Proton-antiproton collisions could produce pairs of the 
gluinos which can decay to bottom quarks and squarks.  When they model 
the Tevatron excess, they find masses for the gluino of 12--16 $GeV$ and 
for the bottom squark of 2--5.5 $GeV$.  While this scenario is not 
standard mSUGRA or gauge mediated SUSY, it is consistent with all 
available constraints from precision measurements at the $Z$, from 
low-energy $e^+ e^-$ experiments, etc.  Recent ALEPH analysis does require 
that the lifetime of the bottom squark be less than a nanosecond.

\section{Partons in the Light Hadrons}

In the hard scattering box, gluons dominate the heavy quark production
process for incident hadrons.  In the case of neutrino production, the 
dominant hard-scattering (via W exchange) occurs off strange quarks in the 
sea.  Thus, fixed-target measurements of charm quark production can tell 
us about the nature and details of the partons in light hadrons.  HERA 
measurements can tell us about the charm content of resolved $\gamma$'s.  
Neutrino and anti-neutrino production of charm quarks can tell us about 
the strange anti-quarks and quarks in target nucleons.  The observed charm 
particle distributions are sensitive to the parton distributions in the 
incident particles, as well as to hadronization effects and the hard 
scattering $\sigma_{c,\overline{c}}$ which produces the charm quarks in 
the first place.

\subsection{Gluon Distributions in Mesons and Nucleons}

From $D$-meson production in Fermilab experiment E769,\cite{e769D} 
it is clear that the 
longitudinal momentum production distributions for incident mesons (pions 
and kaons) are about the same, and much harder than that for incident 
protons.  Given the dominance of gluon-gluon fusion in the production 
process, it is clear from simple kinematics alone that the gluon 
distributions in pions and kaon are about the same, and that the gluons in 
these mesons, carry more momentum than those in baryons.  This 
conclusion should be fairly independent of the details of the hard 
scattering and of the hadronization processes.  
Is it not reassuring that the gluons shared between two quarks typically 
would carry more momentum than the gluons shared among three quarks?


\subsection{Strange Quarks and Antiquarks in the Nucleon Sea}

In neutrino experiments, a strong opposite-sign dimuon signal is observed.
We may expect that one muon (the higher momentum one) comes from the 
charged-current interaction, the other from a charm semileptonic decay.  
The dimuon cross section is dominated by strange quarks; the $d$-quark 
term is small, since ${|V_{cd}|^2}/{|V_{cs}|^2} \sim {1}/{20}$.

The most recent results come from NuTeV, E815 at Fermilab, an experiment 
using a high-energy, sign-selected, quadrupole-triplet, neutrino beam with 
a massive steel detector and from the previous CCFR experiment with the 
same detector.  In the CCFR experiment, the beam is wide-band and 
undifferentiated by sign, thus mixing neutrinos and anti-neutrinos.  The
data analysis requires a comparison of data to Monte Carlo models of 
production and detector.  Parameterization is needed (charm mass, 
fragmentation, and for the sea quark distribution of interest).  
Nevertheless, as summarized in Table \ref{table2}, the strange sea is 
measured to be $\sim 40 \%$ of the non-strange sea.  So far, the analysis 
is only done using a leading-order model.\cite{NuTeV}

\begin{table}
\caption{Strange Sea Quarks as a Fraction of the Non-Strange Sea Quarks}
\centering
\begin{tabular}{|c|c|c|}
\hline   
                       &     from $\nu$      &  from $\overline{\nu}$ \\
\hline
$k = 2s/(u+d)_{sea}$   &   $0.36 \pm 0.05$   &    $0.38 \pm 0.04$ \\
\hline
\end{tabular}
\label{table2}
\end{table}

\subsection{Experimental Evidence on Intrinsic Charm}

If there are strange quarks in the nucleon sea, why not charm quarks?
Initially, significant numbers of charm quarks and antiquarks among the 
partons in the sea was proposed to explain apparently very large forward 
$\Lambda_c$ production at CERN ISR experiments.  These intrinsic charm 
quarks were suggested to carry 1-2\% of the proton momentum.  Such 
intrinsic charm pairs would be co-moving with the valence quarks of the 
parent projectile - making coalescence with them easy, and producing large 
particle/antiparticle asymmetries in the forward direction at low 
$p_t$.\cite{vogtAsym}  So, intrinsic charm might have explained both 
forward charm excesses and, combined with recombination effects, 
observed particle/antiparticle production asymmetries,  However, 
differential cross sections for $J/\psi$ production at high $x_F$ from 
Fermilab experiment E789, limit intrinsic charm to less than 1\% of this 
prediction (corresponding to less than 0.02\% of proton).\cite{e789}  
Furthermore, observed particle/antiparticle production asymmetries in 
hadroproduction appear to be essentially flat with $p_t$,\cite{e791asym}
as in Pythia's modeling of the effect, but not as predicted by intrinsic 
charm models. \cite{e791asym}

\subsection{Intrinsic $k_t$}

So far, we have focused mostly on the longitudinal parton distributions.  
What can we say about the transverse momentum distribution of partons in 
hadrons?  The expression intrinsic ''$k_t$" applies to this initial parton
transverse momentum.  It is cited as accounting for the very large 
final-state charm particle transverse momentum seen in experiments.  
However, the values for intrinsic $k_t$ which result from analyses turn 
out to be unphysically large, 1 or 2 $GeV$ or more, even for particles 
with rest masses below 1 $GeV$.  Clearly, intrinsic $k_t$ is a misnomer 
for something else.  What is it?

Intrinsic $k_t$ is used in models also to soften the back-to-back 
correlations of heavy-quark production, for example of $D$ and anti-$D$ 
production in fixed-target experiments.  The smearing is significantly 
greater in hadroproduction, where there is $k_t$ in both target and 
projectile, than in photoproduction, where there would be significant 
$k_t$ only for the target.  Detailed fits have not been performed, 
however, $k_t$ of about 2 $GeV$ is needed to explain hadroproduction 
smearing,\cite{e791corr} while 1 $GeV$ seems to be about right for 
photoproduction.\cite{e831corr}  See the presentation of Erik Gottschalk 
later in these proceedings.\cite{gottschalk}

\section{Summary}

There is a lot we can learn from heavy-flavor production in the large 
variety of environments to be covered at this Multiparticle Dynamics 
Symposium.  We will see tests of QCD - the processes that matter, and how 
to treat them quantitatively.  We will see fundamental quantities entering 
each process, and use the heavy quark to tag the processes of interest.
On to the data and their interpretation!

\end{document}